\documentstyle[aps,amsfonts,epsfig,multicol]{revtex}

\begin{document}
\title{One dimensional drift-diffusion between two absorbing
boundaries: application to granular segregation}
\author{Z\'en\'o Farkas$^{1,2}$ and Tam\'as F\"ul\"op$^3$}
\address{$^1$Department of Biological Physics, E\"otv\"os University,
Budapest, P\'azm\'any P. Stny 1A, 1117 Hungary\\
$^2$Department of Theoretical Physics, Gerhard-Mercator University, 
D-47048 Duisburg, Germany\\
$^3$Department of Theoretical Physics, E\"otv\"os University,
Budapest, P\'azm\'any P. Stny 1A, 1117 Hungary}
\date{October 11, 2000}
\maketitle


\newcommand\nc[2]{\newcommand#1{#2}}
\nc{\width}{L}
\nc{\rnc[2]}{\renewcommand#1{#2}}
\nc{\ds}{\displaystyle}
\nc{\be}{\begin{equation}}
\nc{\ee}{\end{equation}}
\nc{\bea}{\begin{eqnarray}}
\nc{\eea}{\end{eqnarray}}
\nc{\intlim[2]}{\int\limits_{#1}^{#2}}
\nc{\sumlim[2]}{\sum\limits_{#1}^{#2}}
\nc{\sumvar}{k}
\nc{\f[2]}{\frac{#1}{#2}}
\nc{\mb[1]}{{\mbox{#1}}}
\nc{\idx[1]}{\mb{\scs #1}}
\nc{\mspacea}{\,\,\,\,\,\,}
\nc{\mspaceb}{\,\,\,}
\nc{\scs}{\scriptsize}
\nc{\p}{\partial}
\nc{\dt}{\partial_t}
\nc{\dx}{\partial_x}
\nc{\dxx}{\partial^2_x}
\nc{\scalprod[2]}{\langle #1 | #2 \rangle}
\nc{\dom}{{\mathcal D}}
\nc{\avrg[1]}{\langle #1 \rangle}

\nc{\Hz}{\,\mb{Hz}}
\nc{\mm}{\,\mb{mm}}
\nc{\cm}{\,\mb{cm}}
\nc{\cmps}{\,\frac{\mb{\scs cm}}{\mb{\scs s}}}
\nc{\cmsps}{\,\frac{\mb{\scs cm$^2$}}{\mb{\scs s}}}

\nc{\probdens}{p}
\nc{\pxt}{\probdens(x,t)}
\nc{\current}{j}
\nc{\splitprob}{n}
\nc{\leftside}{\leftarrow}
\nc{\rightside}{\rightarrow}
\nc{\out}{\idx{out}}
\nc{\vel}{v}
\nc{\mfptime}{\tau}
\nc{\diffcoeff}{D}
\nc{\ratio}{\alpha}
\nc{\operator}{F}
\nc{\Doperator}{D_\operator}
\nc{\ef[1]}{\sqrt{\frac{2}{\width}} e^{x/2}\sin\left(\frac{#1\pi}{\width}x\right)}
\nc{\ev[1]}{-\left(\frac{#1\pi}{\width}\right)^2-\frac{1}{4}}
\nc{\evn[1]}{\left(\frac{#1\pi}{\width}\right)^2+1/4}

\newcommand{\qu}{u}
\newcommand{\qul}{{u_1}}
\newcommand{\qur}{{u_2}}
\newcommand{\qq}{q}
\newcommand{\qqq}{\qq'}
\newcommand{\qlambda}{\lambda}
\newcommand{\qcl}{{c_1}}
\newcommand{\qcr}{{c_2}}
\newcommand{\qordo}{{\cal O}}
\newcommand{\qfel}{\frac{1}{2}}
\newcommand{\qie}{{\it i.e.,}}
\newcommand{\qalopt}{\ratio_{\mbox{\scriptsize{optimal}}}}

\newcommand{\bel}[1]{\begin{equation}\label{eq:#1}}


\begin{abstract}

Motivated by a novel method for granular segregation, we analyze
the one dimensional drift-diffusion between two absorbing boundaries.
The time evolution of the probability distribution and the rate of
absorption are given by explicit formulae,
the splitting probability and the mean first passage
time are also calculated.
Applying the results we find optimal parameters for
segregating binary granular mixtures.

\end{abstract}

\begin{multicols}{2}

\section{Introduction}
The diffusion phenomena had been one of the most 
intensively studied fields in statistical physics.
A number of textbooks have been written in this field,
and in many of them the one dimensional case is discussed in detail
\cite{crank,gardiner,risken,kampen}.
Still, according to our knowledge, the special case when the 
diffusing particle is
between two absorbing boundaries, and a constant external force
is applied, has not been completely solved.

The need for the solution of this problem arose when
we investigated the motion of a granular particle in a vertically
vibrated ratchet by means of a 2D computer simulation
(see Fig.\ \ref{fig:ratchet}).
\begin{figure}
\centerline{\epsfig{file=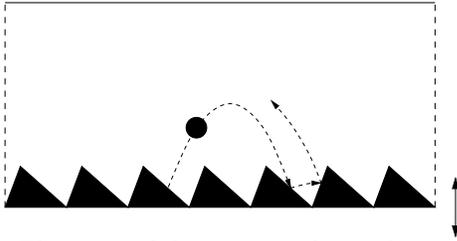,width=0.7\linewidth}}
\caption{The setup of the one particle simulation. The two dimensional
box has an asymmetrical sawtooth shaped base,
which is sinusoidally vibrated in vertical direction
with amplitude $A$ and frequency $f$. The shape of the sawtooth
can be described by three parameters:  width $w$, height $h$, and
asymmetry parameter $a$, which is the ratio of the projection of the
left edge onto the base and $w$. The particle has three parameters:
radius $r$, coefficient of restitution $\varepsilon$ and friction
coefficient $\mu$. The boundary condition can be either periodic,
reflective, or absorbing.}
\label{fig:ratchet}
\end{figure}
For a detailed description of the setup see \cite{farkas}.
We found that the horizontal motion of one particle
can be well approximated as drift-diffusion (see Figures
\ref{fig:driftdiff0.45} and \ref{fig:driftdiff0.6}),
and the parameters of the
diffusion process, i.e. the average velocity and the diffusion constant,
depend on the parameters of the particle and the ratchet \cite{note:segr}.
According to our results, it is possible that the average
velocities of two kinds of
particles in the same system have opposite directions.
If the boundaries are open (i.e.\ absorbing boundary
condition is applied), then this setup is capable of
segregating a binary granular mixture of
these particles provided that the load rate is chosen so that
only a few particles are in the system at one time (in this case
the interaction between the particles can be neglected, and we
do not have the problem that the transport velocity also depends on the
number of particles).
We need the theoretical description to predict e.g.\
the quality of the segregation or the highest possible load rate,
and to further improve the quality
by choosing optimal parameters,
such as where to load the granular mixture in along the box.
The results of the granular binary mixture segregation
will be published elsewhere.

\begin{figure}
\centerline{\epsfig{file=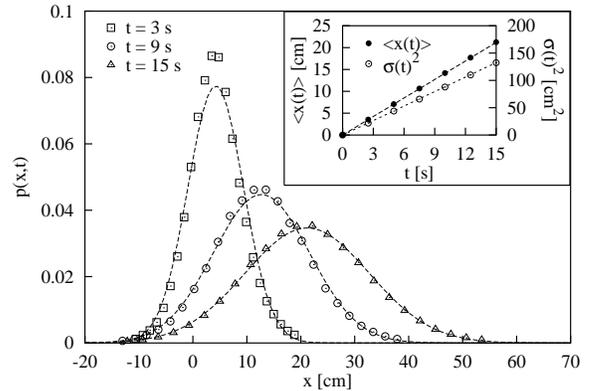,height=0.9\linewidth,angle=270}}
\caption{Simulation result: the probability distribution of
the horizontal position of a particle
which is started from $x=0$ at zero time,
in an infinitely wide system (natural boundary condition).
The dashed curves show the theoretical prediction
$\f{1}{\sqrt{4\pi Dt}}e^{-\f{(x-vt)^2}{4Dt}}$, which is the
probability distribution of a drift-diffusing particle in one dimension.
The velocity $v$ and diffusion constant $D$ are determined by line fitting:
$\avrg{x(t)} = v t$ and $\sigma(t)^2 = 2 D t$
(see inset), where
$\sigma(t)^2 = \avrg{x(t)^2} - \avrg{x(t)}^2$.
The fitted values are $v=1.42 \cmps$ and $D=4.42 \cmsps$.
The parameters of the simulation are the following:
$A = 2 \mm$, $f = 28 \Hz$, $w = 6 \mm$, $h = 8.5 \mm$, $a = 0.07$,
$r = 1.2 \mm$, $\varepsilon = 0.45$, and $\mu = 0.1$.}
\label{fig:driftdiff0.45}
\end{figure}
\begin{figure}
\centerline{\epsfig{file=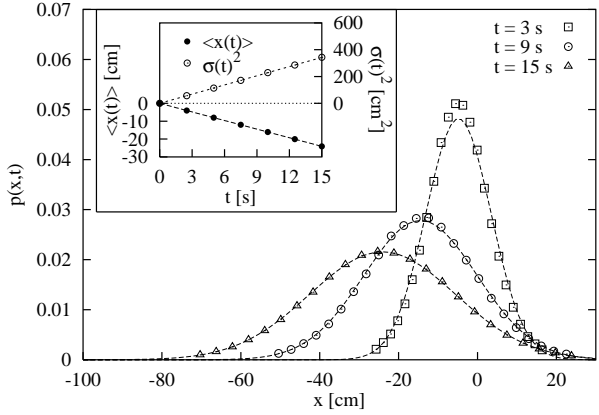,height=0.95\linewidth,angle=270}}
\caption{The same as in Fig.\ \ref{fig:driftdiff0.45}, except for that the
restitution coefficient of the particle is $\varepsilon=0.6$.
The fitted diffusion parameters are $v=-1.60\cmps$ and $D=11.43\cmsps$.
Note that the direction of the velocity is the opposite,
which makes possible the segregation of a granular mixture consisting of
particles with $\varepsilon=0.45$ and $\varepsilon=0.6$.}
\label{fig:driftdiff0.6}
\end{figure}

\section{The diffusion equation with bias}
The dynamics of the diffusing particle is characterized by the
diffusion constant $\diffcoeff$ and mean velocity
$\vel \ne 0$, which is the result of an external field. 
(The results for the $\vel = 0$ case will be presented later.)
The diffusion equation then reads
$
\dt \pxt = \diffcoeff \dxx \pxt - \vel \dx \pxt.
$
The relevant quantities are $\diffcoeff$ and
$\vel$, which define a characteristic time
$t^*=\diffcoeff/\vel^2$ and a characteristic length
$l^*=\diffcoeff/\vel$.
We use $t^*$ and $l^*$ to non-dimensionalize the problem,
all quantities are dimensionless in the rest of this section.
The dimensionless diffusion equation is
\be
\dt \pxt = \dxx \pxt - \dx \pxt
\label{eq:diffusdimless}
\ee
with initial condition
$\probdens(x,0) = \probdens_0(x)$ for $0\le x \le \width$,
and boundary condition
$\probdens(0,t)=\probdens(\width,t)=0$ for $t \ge 0$,
where $\width$ is the system width, and we start at $t=0$.
The absorbing boundary condition is equivalent to that the probability
is zero at the boundaries \cite{risken}.
Let us define operator $\operator=\dxx - \dx$ with
$\dom_\operator = \{\phi(x)|\phi(0)=\phi(\width)=0\}\cap \dom^2$,
where $\dom^2$ is the set of twice differentiable functions.
If $\phi_\lambda \in \dom_\operator$ is an eigenfunction of
$\operator$ with eigenvalue $\lambda$, then
$\exp(\lambda t) \phi_\lambda$ is a solution of (\ref{eq:diffusdimless}).
If we had an orthonormal and complete eigensystem of
$\operator$, we could give the solution of
(\ref{eq:diffusdimless}) immediately.
It is a problem that $\operator$ is not Hermitian,
as far as the usual scalar product
$\int_0^\width \phi_1^*(x) \phi_2(x) dx$
is concerned.
However, $\operator$ is Hermitian if the scalar product
is calculated with kernel function
$\ds e^{-x}$:
$\scalprod{\phi_1}{\phi_2} \equiv \int_0^\width
e^{-x}\phi_1^*(x) \phi_2(x) dx$.
It is useful to write $\operator$ in the form
$\ds \operator=e^x\dx e^{-x} \dx$.
The verification of that $\operator$ is Hermitian with scalar product
$\scalprod{\cdot}{\cdot}$, i.e.\
$\scalprod{\phi_1}{\operator\phi_2} = \scalprod{\operator\phi_1}{\phi_2}$
is straightforward: the effect of
operator $\operator$ can be shifted off from
$\phi_2$ to $\phi_1$ using two consecutive partial integration,
during which the surface terms disappear,
because $\phi_1(x)$ and $\phi_2(x)$ are zero on the borders.
Since $\operator$ is Hermitian with scalar product
$\scalprod{\cdot}{\cdot}$,
it has a complete orthonormal eigenfunction system with real eigenvalues.
It is straightforward to verify that the eigenfuctions are
$\probdens_\sumvar(x) = \ef{\sumvar}$ with eigenvalues
$\lambda_\sumvar = \ev{\sumvar}\mspacea$ $(\sumvar \in \mathbb{N}^+).$
The solution of (\ref{eq:diffusdimless}), 
if $\probdens^0(x)$ is the initial probability distribution, is
$\probdens(x,t)=\sumlim{\sumvar=1}{\infty}
\scalprod{\probdens^0}{\probdens_\sumvar} \probdens_\sumvar(x)
e^{\lambda_\sumvar t}$.
In the rest of the paper we investigate the case when the initial
distribution is a Dirac-delta function at
$\ratio \width$ ($0<\ratio<1$). The weights are then
$\scalprod{\delta(x-\ratio \width)}{\probdens_\sumvar} =
\sqrt{\frac{2}{\width}} e^{-\frac{\ratio \width}{2}}
\sin(\sumvar\pi \ratio)$,
and the probability distribution is given by
\bea
\probdens(x,t) &=& \frac{2}{\width} e^{\frac{x-\ratio\width}{2}}
\sumlim{\sumvar=1}{\infty} \sin(\sumvar\pi \ratio)
\sin\left(\frac{\sumvar\pi}{\width}x\right)
e^{-\left[\left(\frac{\sumvar\pi}{\width}\right)^2
+\frac{1}{4}\right] t}\nonumber\\
&=& \frac{1}{2\width} e^{\frac{x-\ratio\width}{2}-\frac{t}{4}}
\bigg\{\vartheta_3\bigg[\bigg(\ratio-\frac{x}{\width}\bigg)\frac{\pi}{2},
z(t)\bigg]\nonumber\\
& &- \vartheta_3\bigg[\bigg(\ratio+\frac{x}{\width}\bigg)\frac{\pi}{2},
z(t)\bigg]\bigg\},
\eea
where we introduce the notation
$z(t) = e^{-\f{\pi^2}{\width^2}t}$
and the theta function
$ \vartheta_3(r,q) = 1+2 \sumlim{\sumvar=1}{\infty}\cos(2r\sumvar)
q^{\sumvar^2}$
to obtain a closed form \cite{abramowitz}.
An important quantity is the rate of absorption, i.e.\ the
probability current at the borders.
Using the notation
$\vartheta_3'(r,q)\equiv \partial_{\tilde r}
\vartheta_3(\tilde r,q)|_{\tilde r = r}$,
the probability current
$\current(x,t)= -\dx \probdens(x,t) + \probdens(x,t)$
at the right and left end is
\bea
\current_\rightside(t) &\equiv& \ds \current(\width,t) =
\f{\pi}{2\width^2} e^{\f{(1-\ratio)\width}{2}-\f{t}{4}}
\vartheta_3'\left(\f{\pi[\ratio+1]}{2}, z(t)\right),\nonumber\\
\current_\leftside(t) &\equiv& \ds -\current(0,t) =
-\f{\pi}{2\width^2} e^{-\f{\ratio\width}{2}-\f{t}{4}}
\vartheta_3'\left(\f{\pi \ratio}{2}, z(t)\right).
\label{eq:current}
\eea
With the minus sign in its definition, $\current_\leftside(t) \ge 0$.
\begin{figure}
\centerline{\epsfig{file=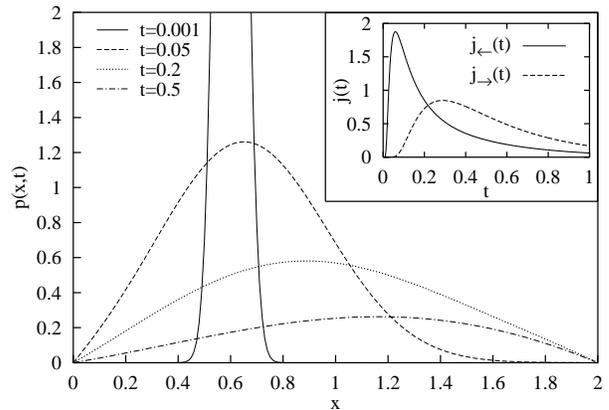,height=0.95\linewidth,angle=270}}
\caption{The time evolution of the probability distribution, when
$\width=2$ and $\ratio=0.3$.
Note the difference between the probability distribution here and in
Figs.\ \ref{fig:driftdiff0.45} and \ref{fig:driftdiff0.6}, which is
due to the different boundary condition.
Inset: the probability current at the left and right boundary.}
\label{fig:current}
\end{figure}
The splitting probablility, i.e.\ the
probability that the particle is absorbed finally by the
left or the right boundary:
$\splitprob_\leftside = \int_0^\infty \current_\leftside(t) dt$, and
$\splitprob_\rightside = \int_0^\infty \current_\rightside(t) dt$,
and obviously $\splitprob_\leftside + \splitprob_\rightside = 1$.
Using the integral formula
\be
\intlim{0}{\infty} e^{-a s} \vartheta_3'\left(\left[\f{x}{l}+1\right]
\f{\pi}{2}, e^{-\f{\pi^2}{l^2}s}\right) ds =
\f{2 l^2}{\pi} \f{\sinh(x\sqrt{a})}{\sinh(l\sqrt{a})}
\label{eq:intthetaprime}
\ee
which holds for $|x| < l$ \cite{erdelyi}, we get
\be
\splitprob_\leftside = \frac{e^{-\ratio\width}-e^{-\width}}{1-e^{-\width}}
\mspacea \mb{and} \mspacea
\splitprob_\rightside = \frac{1-e^{-\ratio\width}}{1-e^{-\width}}.
\label{eq:splitprob}
\ee

Another important quantity is the {\em mean first passage time},
i.e.\ the average time it takes for the particle
to be absorbed by any of the boundaries.
The probability distribution of this time is
just the total probability current at the boundaries:
$\current_\out = \current_\leftside + \current_\rightside$.
The mean first passage time is then
$\mfptime = \int_0^\infty t \current_\out(t) dt$.
The calculation of $\mfptime$ is straightforward using the derivative
of (\ref{eq:intthetaprime}) w.r.t.\ $a$, and the result is
\be
\mfptime = \width (\splitprob_\rightside - \ratio).
\label{eq:mfptime}
\ee

There is another way to calculate $\splitprob_\rightside$ and $\mfptime$,
in which there is no need for an explicit formula for the time
dependent probability distribution or the current at the
borders \cite{gardiner}.
An ordinary differential equation can be written for
$\splitprob_\rightside$, which, using our notation, is
$
-\width \p_\ratio \splitprob_\rightside
+ \p^2_\ratio \splitprob_\rightside = 0
$
with boundary condition $\splitprob_\rightside|_{\ratio=0} = 0$ and
$\splitprob_\rightside|_{\ratio=1} = 1$.
By direct substitution one can verify that (\ref{eq:splitprob})
is the solution.
A similar equation can be written for the mean first passage time:
$
-\width \p_\ratio \mfptime + \p^2_\ratio \mfptime = -L^2
$
with boundary condition $\mfptime|_{\ratio=0} = 0$ and
$\mfptime|_{\ratio=1} = 0$. It is easy to check that
(\ref{eq:mfptime}) is the solution for this equation.

\section{Discussion of the results}
First we summarize our results with dimensionalized parameters.
The probability density is
\bea
\probdens(x,t) &=& \frac{\vel}{2\diffcoeff\width}
e^{-\f{\vel(2[\ratio\width-x]+\vel t)}{4\diffcoeff}}
\bigg\{
\vartheta_3\bigg(\bigg[\ratio-\f{x}{\width}\bigg]\f{\pi}{2},
\tilde z(t)\bigg)\nonumber\\
&&- \vartheta_3\bigg(\bigg[\ratio+\f{x}{\width}\bigg]\frac{\pi}{2},
\tilde z(t)\bigg)\bigg\},
\label{eq:probdensdim}
\eea
with the notation
$\tilde z(t) \equiv e^{-\pi^2 \diffcoeff t/\width^2}$.
The currents at the right and left borders are
\bea
\current_\rightside(t) &=& \ds \f{\pi\vel^2}{2\diffcoeff\width^2}
e^{-\f{\vel(2[\ratio-1]\width+\vel t)}{4\diffcoeff}}
\vartheta_3'\left(\f{[\ratio-1]\pi}{2}, 
\tilde z(t)\right),\nonumber\\
\current_\leftside(t) &=& \ds -\f{\pi\vel^2}{2\diffcoeff\width^2}
e^{-\f{\vel(2\ratio\width+\vel t)}{4\diffcoeff}}
\vartheta_3'\left(\f{\ratio\pi}{2},
\tilde z(t)\right).
\label{eq:currentdim}
\eea
The splitting probability is
$\ds n_\rightside = \frac{1-e^{-\ratio\width\vel/\diffcoeff}}
{1-e^{-\width\vel/\diffcoeff}}$,
and  the mean first passage time is
$\mfptime = \width(n_\rightside - \ratio)/\vel$.
In the $\vel = 0$ case (\ref{eq:probdensdim}) and (\ref{eq:currentdim})
are valid with $\vel=0$ substitution,
while for the splitting probability and the mean first passage time
we can take the $\vel \rightarrow 0$ limit
and obtain $\splitprob_\rightside = \ratio$ and 
$\mfptime = \ratio (1-\ratio) L^2/(2D)$.

Since the splitting probability is the most important quantity as
far as the segregation is concerned, we analyze it in detail.
Its dependence on all four parameters $v$, $D$, $\alpha$, and $L$ can be
seen in Fig.\ \ref{fig:splitprob}.
It can be easily understood that with larger (positive) velocity
or smaller diffusion constant the right arriving probability 
is closer to $1$. It is also trivial that if the starting position
is closer to the right end, then $\splitprob_\rightside$ is larger.
However, to understand the dependence on the system width, we have to
recall that in the case of drift-diffusion with natural boundary condition
(i.e.\ when the system is infinitely wide)
the expectation value of the position is proportional to the time, while
the dispersion is proportional only to the square root of time.
Although this is not exactly true in the case of absorbing boundaries,
the tendencies remain the same,
therefore, in case of $\vel > 0$,  the larger the system width is,
the less part of the probability is absorbed by the left boundary,
and as a consequence, the rest is absorbed by the right boundary.

\begin{figure}
\centerline{\epsfig{file=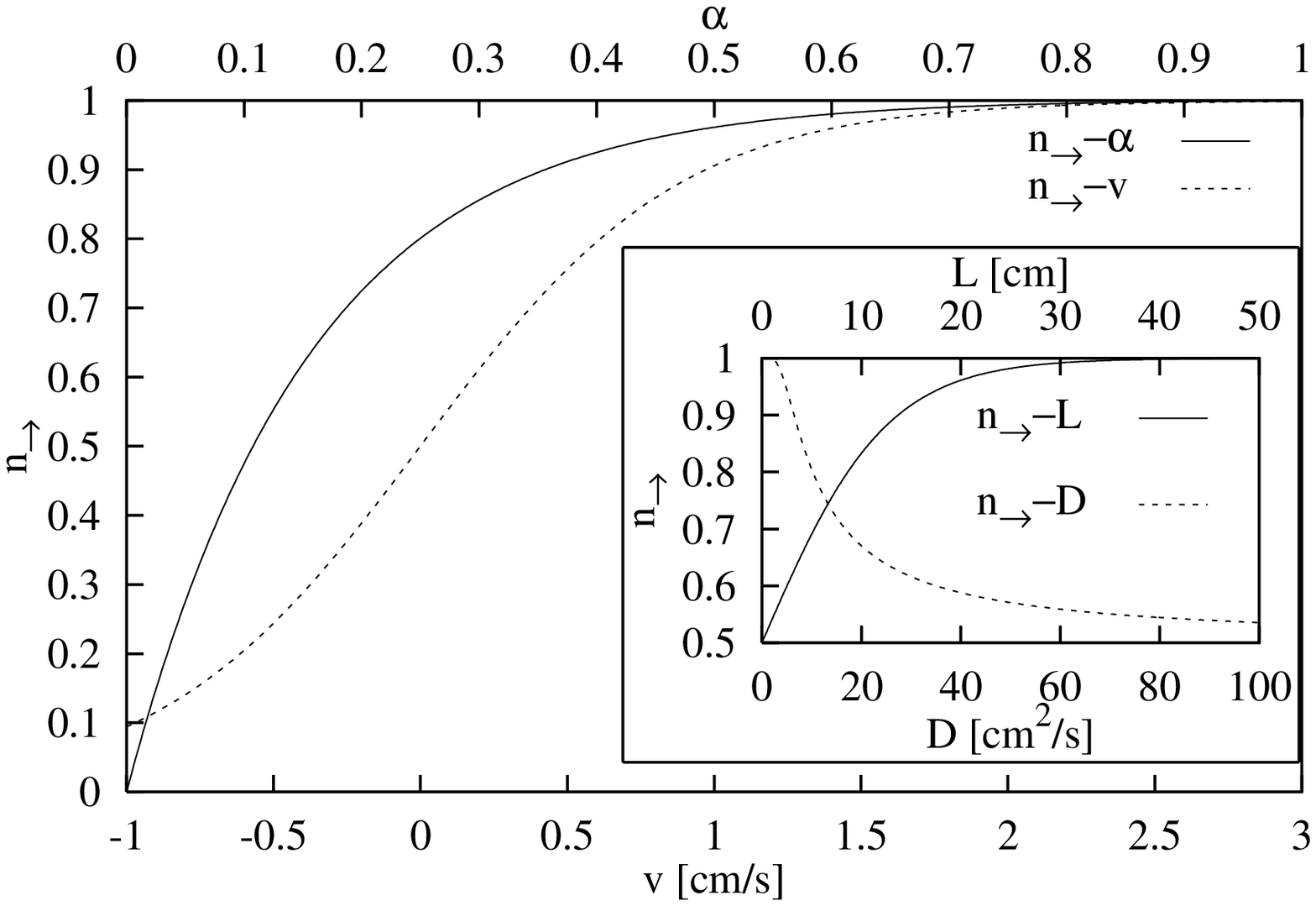,height=0.95\linewidth,angle=270}}
\caption{Dependence of the splitting probability
$\splitprob_\rightside$ on all the parameters.
The default values for the parameters are
$\vel=1.42 \cmps$,
$\diffcoeff=4.42 \cmsps$ (the fitted values in Fig.\ \ref{fig:driftdiff0.45}),
$\ratio=0.5$, and $\width=20\cm$.}
\label{fig:splitprob}
\end{figure}

\section{An application: Segregation}

As mentioned in the Introduction, our motivation for 
the above calculations has been the application for segregating binary 
granular mixtures. For this purpose, now let us investigate the 
segregation properties of the system.

We begin with an illustrative example, the case of symmetric initial 
condition, $ \ratio = 1/2 $. Denoting $\vel / \diffcoeff$ by $\qu$, the 
splitting probabilities are 
    \bel{aa}
    \splitprob_\leftside = \frac{1}{ 1 + e^{   \qu \width / 2 } } , \mspacea
    \splitprob_\rightside = \frac{1}{ 1 + e^{ - \qu \width / 2 } } .
    \ee
For definiteness, let us have a particle with $\qu > 0$, for example. 
For a small $\width$, both probabilities are around $1/2$:
    \bel{ab}
    \splitprob_\leftside \approx 1/2 - \qu \width / 8 , \mspacea
    \splitprob_\rightside \approx 1/2 + \qu \width / 8 .
    \ee
 For large $L$s,
    \bel{ac}
    \splitprob_\leftside \approx     e^{ - \qu \width / 2 } , \mspacea
    \splitprob_\rightside \approx 1 - e^{ - \qu \width / 2 } ,
    \ee
they tend to zero and one, respectively, in an exponential way in $\width$.
We can see that, for a small $\width$,
it is the diffusion, the left-right
symmetric effect, that determines the left and right probabilities. On the
other side, for large $\width$s, the drift becomes the dominating effect, it
drives the particle to the direction of $\qu$ with a probability that
differs from $1$ only by an exponentially small amount. This latter
phenomenon makes it possible to use the system for segregation.

To study the segregation properties in more detail, now let us have two 
types of particles, with $ \qul < 0 < \qur $.
We neglect the interaction between particles,
which is a good approximation at low particle numbers.
The expectation is that 
the particles with $\qul$ will tend to move to the left end and the 
others to the right end. Therefore, we can characterize the quality of 
the segregation with the quantity
    \bel{ad}
    \qq = \splitprob_\leftside (\qul) + \splitprob_\rightside (\qur) .
    \ee
Let us put the question whether, for a given $\width$, it is possible to 
choose an optimal $\ratio$, \qie where $\qq$ is
maximal \cite{note:qchoice}.

Investigation of the function $\qq(\ratio)$ shows that it indeed 
has one maximum, at
    \bel{ae}
    \qalopt = 1 - \frac{ \ln \frac{ \qul (e^{ \qur \width } - 1) }{ \qur 
    (e^{\qul \width} - 1) } }{ (\qur - \qul) \width } .
    \ee
For an illustration of the typical dependence of $\qq$ on $\ratio$, see 
Fig.\ \ref{fig:optratio}.
\begin{figure}
\centerline{\epsfig{file=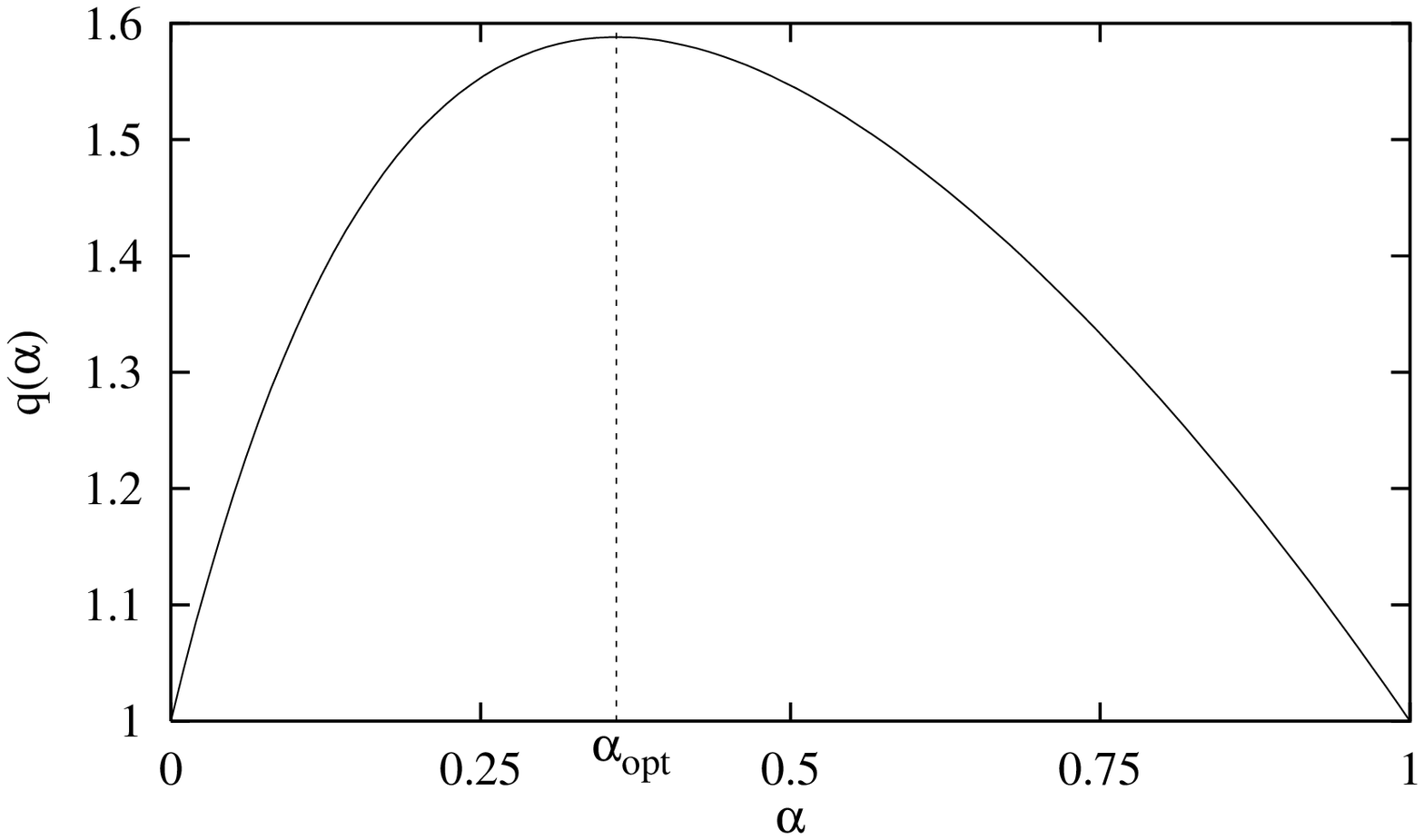,height=0.95\linewidth,angle=270}}
\caption{Dependence of $\qq$ on $\ratio$, for $ \qul \width = - 1 $ 
and $ \qur \width = 5 $.}
\label{fig:optratio}
\end{figure}

For practical purposes, we are interested in large $\width$s. In this 
asymptotic region, with the notation $ \qlambda = \qur / |\qul| $,
    \bel{af}
    \qalopt \approx
    \frac{1}{1 + \qlambda} + \frac{ \ln \qlambda }{1 +\qlambda} \;
    \frac{1}{ |\qul| \width } \, = \,
    \frac{1}{1 + \qlambda} + \qordo \left( \frac{1}{L} \right) .
    \ee
Consequently,
    \bea
    \splitprob_\leftside (\qul) &\approx& 1 - \qcl e^{ - \frac{\qlambda}{1 + \qlambda}
    |\qul \width| },\nonumber\\
    \splitprob_\rightside (\qur) &\approx& 1 - \qcr
    e^{ - \frac{1}{1 + \qlambda} |\qur \width| } ,
    \eea
where $ \qcl = \qlambda^{ \frac{1}{1 + \qlambda} } $ and $ \qcr =
\qlambda^{ \frac{- \qlambda}{1 + \qlambda} } $. Therefore, we can see
that, for large $\width$s, it is possible to choose such an optimal $\ratio$
that both segregation probabilities are exponentially close to $1$ (as
functions of $\width$), the quality of the segregation is exponentially close 
to the perfect.

\section{Summary}
We presented a complete solution to 
the problem of one dimensional
drift-diffusion (with constant external force) between
two absorbing boundaries: the probability distribution, the
probability current at the boundaries (i.e.\ the rate of absorption),
the splitting probability and the mean first passage time were
calculated.
The results were applied to predict
the quality of granular segregation in a vertically vibrated ratchet.
We found that if the components have opposite drift velocities,
the quality of the segregation of a binary mixture increases rapidly
with increasing system width, and as a limiting case,
perfect segregation can be achieved.
Furthermore, when the system width is fixed, we found the place
where the granular mixture should be loaded into the system to
obtain the best segregation quality.

\section{Acknowledgments}
We thank Tam\'as Vicsek and Dietrich E. Wolf for their helpful
advices.

\end{multicols}


\begin{thebibliography}{x}
\bibitem{note:segr} The detailed analysis of the one particle motion
and the segregation of binary granular mixtures using this method
will be published elsewhere.
\bibitem{note:qchoice} Another natural choice to describe the quality of the 
segregation is $ \qqq = \splitprob_\leftside (\qul) \cdot
\splitprob_\rightside (\qur) $. Since the two 
functions, $\qq(\ratio)$ and $\qqq(\ratio)$, prove to differ only very 
slightly and lead to almost identical $\qalopt$, the two quantities are 
practically equivalent choices.
\end{thebibliography}
\end{document}